\documentclass[aps,pra,twocolumn,showpacs,superscriptaddress]{revtex4-2}  

\usepackage{physics}
\usepackage{graphicx}  
\usepackage{dcolumn}   
\usepackage{bm,amssymb,amsmath,dsfont}   
\usepackage[colorlinks=true,breaklinks=true,allcolors=blue]{hyperref}
\usepackage{url}
\usepackage{txfonts}
\usepackage{bbold}
\usepackage{xr}

\usepackage{dsfont}

\usepackage{graphics}
\usepackage{pstricks}
\usepackage{tikz}
\usetikzlibrary{decorations.markings}
\usepackage{color}
\usepackage{xcolor}
\graphicspath{{Figures/}}

\usepackage{tabularx}
\usepackage{multirow}
\usepackage{array}
\newcolumntype{L}[1]{>{\raggedright\let\newline\\\arraybackslash\hspace{0pt}}m{#1}}
\newcolumntype{C}[1]{>{\centering\let\newline\\\arraybackslash\hspace{0pt}}m{#1}}
\newcolumntype{R}[1]{>{\raggedleft\let\newline\\\arraybackslash\hspace{0pt}}m{#1}}

\newcommand{\e}{\mathrm{e}}
\newcommand{\id}{\mathrm{d}}

\newcommand{\eq}[1]{(\ref{eq:#1})}
\newcommand{\Eq}[1]{Eq.\,\eq{#1}}

\newcommand{\Fig}[1]{Fig.~\ref{fig:#1}}
\newcommand{\fig}[1]{\ref{fig:#1}}

\newcommand{\App}[1]{App.~\ref{app:#1}}

\definecolor{applegreen}{rgb}{0.55, 0.71, 0.0}
\definecolor{byzantine}{rgb}{0.74, 0.2, 0.64}

\newcommand{\cred}[1]{{\color{red}{#1}}}
\newcommand{\cblu}[1]{{\color{blue}{#1}}}

\newcommand{\tbd}[1]{\cred{#1}}
\renewcommand{\tbd}[1]{}

\newcommand{\tbdso}[1]{\cblu{#1}}
\renewcommand{\tbdso}[1]{}

\newcommand{\IntermediateStep}[1]{&\cred{\textrm{\ (($===$ intermediate calc. steps $==>$))}}\nonumber\\ #1  
&\cred{\textrm{(($<=====================$))}}\nonumber\\}
\renewcommand{\IntermediateStep}[1]{}

\makeatletter
\let\cat@comma@active\@empty
\makeatother

\begin{document}


\title{Ab initio Complex Langevin computation of the roton gap for a dipolar Bose condensate}
\author{Philipp Heinen}
\affiliation{Kirchhoff-Institut f\"ur Physik,
             Universit\"at Heidelberg,
             Im~Neuenheimer~Feld~227,
             69120~Heidelberg, Germany}
\author{Wyatt Kirkby}
 \affiliation{Kirchhoff-Institut f\"ur Physik,
             Universit\"at Heidelberg,
             Im~Neuenheimer~Feld~227,
             69120~Heidelberg, Germany}
\affiliation{Physikalisches Institut,
	           Universit\"at Heidelberg,
	           Im~Neuenheimer~Feld~226,
	           69120~Heidelberg, Germany}
\author{Lauriane Chomaz}
\affiliation{Physikalisches Institut,
	           Universit\"at Heidelberg,
	           Im~Neuenheimer~Feld~226,
	           69120~Heidelberg, Germany}
\author{Thomas Gasenzer}
\affiliation{Kirchhoff-Institut f\"ur Physik,
             Universit\"at Heidelberg,
             Im~Neuenheimer~Feld~227,
             69120~Heidelberg, Germany}
\affiliation{Institut f\"{u}r Theoretische Physik,
		      Universit\"{a}t Heidelberg, 
		      Philosophenweg 16, 
		      69120 Heidelberg, Germany}


\begin{abstract}
We compute from first principles the dispersion relation of a finite-temperature dipolar Bose gas near the roton instability. Using a field-theoretic Complex Langevin (CL) approach, we access experimentally relevant atom numbers. We determine the roton excitation gap as a function of the s-wave scattering length under the conditions of the experimental study by Petter et al.~[Phys.~Rev.~Lett.~122, 183401 (2019)] using erbium atoms. Our results show good agreement with experimental measurements close to the instability, in contrast to predictions of the standard approximate field theory based on the extended Gross-Pitaevskii equation (eGPE). While the eGPE agrees with our results far from the instability, we find that it quantitatively fails as the roton softens, where the CL result lies between the GPE and eGPE predictions. This deviation reveals limitations of the (extended) mean-field description and establishes Complex Langevin as a powerful tool for quantitatively accurate simulations of many-body quantum systems where quantum fluctuations and temperature play an essential role.
\end{abstract}

\pacs{%
}

\maketitle

\section{Introduction}
Ultracold Bose gases with strong magnetic dipolar interactions exhibit a range of phenomena that are absent from non-dipolar systems and have thus been the subject of intensive experimental \cite{Chomaz:2022cgi,Griesmaier:2005,Lu:2011,Aikawa:2012,Kadau:2016,Igor:2016,Chomaz:2016,Chomaz:2018,Petter:2019,Tanzi:2019,Boettcher:2019,Chomaz:2019,klaus2022observation,Schmitt:2016} as well as theoretical \cite{Baranov:2008, Baranov2012,ODell2003,Santos:2003,Schuetzhold:2006,glaum2007critical,Lima:2011,Lima:2012,bisset2012finite,Lu2015sds,Waechtler:2016,Zhang2019,roccuzzo2020rotating,Smith:2023} interest. A striking phenomenon is the occurrence of a rotonic feature in the excitation spectrum -- i.e.~the dispersion relation develops a local minimum at a non-zero momentum -- when the dipolar interaction becomes sufficiently strong compared to the contact interaction \cite{Santos:2003,ODell2003,Chomaz:2018}. Rotons had been long known in the context of strongly interacting superfluid helium \cite{Pitaevskii:2016} but do not occur in weakly interacting non-dipolar Bose gases. Dipolar gases thus provide a tunable and weakly interacting platform to study roton physics and related many-body phenomena.

Remarkably, roton softening connects to a tendency towards crystallization, possibly leading to supersolidity~\cite{Feynman1954,Nozieres2004itr}. In a dipolar gas, the roton may be fully softened by fine tuning, e.g., the contact interaction strength~\cite{Santos:2003,Chomaz:2018,Bohn:2009, Lasinio:2013,alana2023supersolid}, leading to a mean-field instability. Yet, experiments have shown that collapse is avoided~\cite{Kadau:2016} thanks to beyond-mean-field effects~\cite{Igor:2016,Chomaz:2016,Waechtler:2016}, resulting in stabilized crystallized ground states, and in particular of supersolids~\cite{Tanzi:2019,Boettcher:2019,Chomaz:2019}. 

Dilute dipolar Bose gases are typically well described by the mean-field Gross-Pitaevskii equation (GPE)~\cite{Pitaevskii:2016}, or more accurately an extended GPE (eGPE)~\cite{Waechtler:2016,Chomaz:2016,Bisset:2016} that additionally includes quantum fluctuations at leading order via the so-called Lee-Huang-Yang (LHY) term~\cite{Schuetzhold:2006,Lima:2011,Lima:2012}. 
While the LHY term captures crucial effects of quantum fluctuations, the eGPE approach remains approximate and not fully self-consistent. 
Recent measurements of the roton gap have highlighted quantitative discrepancies with eGPE predictions~\cite{Petter:2019}, and the role of finite temperature in both the roton softening and the transition to a supersolid state remains an open question~\cite{Sanchez2023heating,Bombin2025creating,LiangJun2025accessing}. This calls for theoretical approaches that accurately include quantum and thermal fluctuations, particularly for large atom numbers.

Path integral Monte Carlo (PIMC) provides an exact ab initio quantum simulation method for interacting bosons \cite{pollock1984simulation,ceperley1986path,pollock1987path,nho2005bose,Cinti:2017,Bombin2019berezinskii,Bombin2024Quantum}. Yet, as a method based on the $N$-particle formulation of quantum mechanics, it is so far limited to relatively small particle numbers compared to the experimentally realistic cases of $10^4$ to $10^5$ atoms. An alternative is the Complex Langevin (CL) algorithm that simulates the field-theoretic path integral in an extended configuration space \cite{sexty2014simulating,kogut2019applying,ito2020complex,rammelmueller2017surmounting,rammelmuller2018finite,Attanasio2023a.PhysRevA.109.033305,Hayata:2015,Attanasio:2020,Delaney:2020, Heinen:2022,Heinen:2023wtt.PhysRevA.108.053311,Heinen2024a}. In this framework, particle number is encoded in the magnitude of the bosonic field $\psi$ and can thus be chosen arbitrarily, while retaining exactness within numerical accuracy. 
The CL approach has a long history in lattice quantum chromodynamics simulations~\cite{sexty2014simulating,kogut2019applying,ito2020complex}, and has more recently been applied to ultracold gases~\cite{rammelmueller2017surmounting,rammelmuller2018finite,Attanasio2023a.PhysRevA.109.033305}, where it reliably describes Bose gases in thermal equilibrium, below and above the superfluid transition~\cite{Hayata:2015, Attanasio:2020,Delaney:2020, Heinen:2022,Heinen:2023wtt.PhysRevA.108.053311,Heinen2024a}. However, it has so far been limited to contact-interacting gases, and lacks quantitative comparison with experiments.

In this work, we apply the CL method to trapped dipolar Bose gases in the mean-field stable regime close to the roton instability at finite temperature. Focussing on the parameters from Ref.~\cite{Petter:2019}, where discrepancies between mean-field theories including and excluding the LHY correction were observed, we compare our CL simulations to both experiments and mean field theories.  

\section{System}
A one-component dipolar Bose gas with dipoles polarized along the $z$-direction, in thermal equilibrium, is described by the following action:
\begin{align}
    \label{eq:action}
    S[\psi,\bar{\psi}]=\int \limits_0^{\hbar\beta} \mathrm{d}\tau\int \mathrm{d}^3r\, \Big\{\hbar\bar{\psi}\partial_\tau\psi
    +\mathcal{H}\left[\psi,\bar{\psi}\right]\Big\}\,,
\end{align}
with Hamiltonian density,
\begin{align}
\label{eq:Hamiltonian}
    \nonumber\mathcal{H}\left[\psi,\bar{\psi}\right]
    &= \frac{\hbar^2}{2m}\nabla\bar{\psi}\cdot\nabla\psi
    -(\mu-V_\text{ext})\,\bar{\psi}\psi
    \\
    &\quad +\frac{g}{2}\left(\bar{\psi}\psi\right)^2+\mathcal{H}_\mathrm{dip}\left[\psi,\bar{\psi}\right]\,,
\end{align}
where the dipolar interaction term is defined as
\begin{align}
    \nonumber\mathcal{H}_\mathrm{dip}\left[\psi,\bar{\psi}\right](\mathbf{r})
    &=\frac{1}{2}\int \id^3r'\,\bar{\psi}(\mathbf{r}')\psi(\mathbf{r}')\\
    &\quad \times\frac{C_\mathrm{dd}}{4\pi}\frac{1-3(z-z')^2/|\mathbf{r}-\mathbf{r}'|^2}{|\mathbf{r}-\mathbf{r}'|^3}\bar{\psi}(\mathbf{r})\psi(\mathbf{r})\,,
\end{align}
Here $m$ denotes the atomic mass, $\mu$ is the chemical potential and $\beta=1/k_\mathrm{B}T$ is the inverse temperature.  $V_\text{ext}(\mathbf{r})=\frac{1}{2}m(\omega_x^2 x^2+\omega_y^2 y^2+\omega_z^2 z^2)$ is the harmonic external trapping potential, with angular frequencies $\omega_{x,y,z}$ in the three spatial directions. The coupling constant $g$ indicates the strength of the contact interaction between the atoms and is at first order related to the s-wave scattering length $a$ by $g=4\pi\hbar^2a/m$ (cf.~the discussion below and \App{Born_series} for higher-order corrections). The coupling term $C_\mathrm{dd}$ encodes the strength of the dipolar interaction. We introduce the dipolar length $a_\mathrm{dd}\equiv mC_\mathrm{dd}/12\pi\hbar^2$ as well as the ratio $\varepsilon_\mathrm{dd}\equiv a_\mathrm{dd}/a$, which measures the relative strength of contact and dipolar interactions. 
The fields $\psi$ and $\bar{\psi}$ are functions of both position $\mathbf{r}$ and imaginary time $\tau$, i.e. $\psi(\tau, \mathbf{r})$ and $\bar{\psi}(\tau, \mathbf{r})$, and fulfill the bosonic boundary condition $\psi(\tau=0, \mathbf{r})=\psi(\tau=\hbar\beta, \mathbf{r})$. %

\section{Complex Langevin}
Within a bosonic quantum field theory described by an action $S[\psi,\bar{\psi}]$, observables $\mathcal{O}[\psi,\bar{\psi}]$ can be computed from the Feynman path integral as
\begin{align}
    \label{eq:pi}
    \langle\mathcal{O}\rangle=Z^{-1}\int \mathcal{D}\psi\mathcal{D}\bar{\psi}\,\exp\left(-S[\psi,\bar{\psi}]/\hbar\right)\,\mathcal{O}[\psi,\bar{\psi}]\,,
\end{align}
with the partition function as normalization, 
\begin{align}
    Z=\int \mathcal{D}\psi\mathcal{D}\bar{\psi}\,\exp\left(-S[\psi,\bar{\psi}]/\hbar\right)\,.
\end{align}
Under the path integral, the two fields $\psi$ and $\bar{\psi}$ are the complex conjugates of each other, $\bar{\psi}=\psi^*$. For a real action $S[\psi,\bar{\psi}]\in\mathbb{R}$, the path integral \eq{pi} can be computed numerically by standard Monte Carlo (MC) methods, such as the Metropolis-Hastings algorithm. However, the action \eq{action} of non-relativistic bosons contains a complex term $\bar{\psi}\partial_\tau\psi$, which renders the application of such MC approaches unfeasible, as it leads to the so-called sign problem \cite{henelius2000a.prb61.1102,lombardo2007a.mpla22.457,alford2010mitigating}.

An approach to overcome the sign problem is the CL method \cite{parisi1983complex,berger2021a.prep892.1}, also termed Stochastic Quantization.  
It consists in rewriting the path integral as a Langevin-type stochastic differential equation in a fictitious time $\vartheta$, of the form,
\begin{align}
    \label{eq:cl_eq12}
    \frac{\partial\psi}{\partial\vartheta}
    &=-\frac{1}{\hbar}\frac{\delta S}{\delta \bar{\psi}}+\eta(\vartheta)
    \,,\quad
    \frac{\partial\bar{\psi}}{\partial\vartheta}
    =-\frac{1}{\hbar}\frac{\delta S}{\delta \psi}+\eta(\vartheta)^*
    \,,
\end{align}
where the complex noise fulfills $\langle\eta\rangle=\langle\eta^*\rangle=\langle\eta^2\rangle=\langle{\eta^*}^2\rangle=0$ and $\langle\eta(\vartheta)\eta(\vartheta')^*\rangle=2\delta(\vartheta-\vartheta')$, and observables are computed as long-time averages, i.e.,
\begin{align}
    \langle\mathcal{O}\rangle
    =\lim_{\Theta\to\infty}\frac{1}{\Theta}\int \limits_{0}^\Theta \id\vartheta\, \mathcal{O}[\psi(\vartheta),\bar{\psi}(\vartheta)]\,,
\end{align} 
in terms of the fields $\psi$ and $\bar{\psi}$, which, besides $\mathbf{r}$ and $\tau$, thus acquire a third dependency on the fictitious time $\vartheta$.

For $S\in\mathbb{R}$, Eqs.~\eq{cl_eq12} are complex conjugates of one another, and $\psi$ and $\bar{\psi}$ remain complex conjugates, $\bar{\psi}=\psi^*$, as in the original path integral. The advantage of reformulating in terms of stochastic differential equations is that we may also consider complex actions, $S\in\mathbb{C}$. Then, $\psi$ and $\bar{\psi}$ will no longer be the complex conjugates, but rather independent complex fields, i.e., $\bar{\psi}\neq\psi^*$. The doubling of degrees of freedom is the cost of avoiding sampling from a complex-valued quasi-probability density, cf. \App{CLangevin} for details.

\begin{figure}
	\includegraphics[width=1.0\columnwidth]{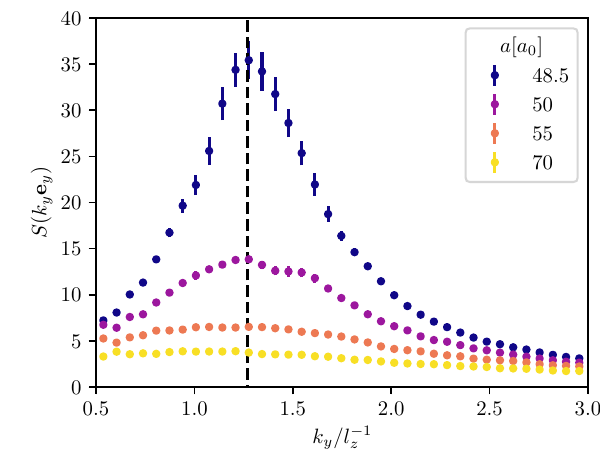}
	\caption{Static structure factor $S(k\mathbf{e}_y)$ as obtained from the CL simulations for a condensate of $N=2.4\times 10^{4}$ Er atoms at $T=50\,\text{nK}$ in a harmonic trap with frequencies $\omega_{x,y,z} = 2\pi \times (261, 27, 256)\,\text{Hz}$ and for scattering length $a$ as given in the legend, in units of the Bohr radius. The dashed black line marks 
    $k_\text{rot}=1.27\, l_z^{-1}$, $l_z=\sqrt{\hbar/m\omega_z}$.
	}
	\label{fig:Sk}
\end{figure}

\section{Simulations}
We consider a gas of erbium atoms in a trap of frequencies $\omega_{x,y,z} = 2\pi \times (261, 27, 256)\,\text{Hz}$ as in Ref.~\cite{Petter:2019}. Here, $m=166 \,\mathrm{u}$, $a_\mathrm{dd}=65.5\,a_0$, and $a$ varies within $a\approx 50\dots80\, a_0$ with $\mathrm{u}$ the atomic mass unit and $a_0$ the Bohr radius. 
The temperature is set to the experimentally relevant value of $T=50\,\text{nK}$, except when comparing temperature dependencies in \Fig{omegak} (b). A total particle number of $N=2.4\times 10^{4}$ or $N=4.8\times 10^{4}$ is used, as relevant for the data of Ref.~\cite{Petter:2019}, see later discussion. 
As the simulations are grand-canonical, we adjust the chemical potential $\mu$ to give a total particle number that deviates by less than $1\%$ from  the targeted $N$ values. We discretize the action \eq{action} on a periodic lattice twice as large as the atom cloud (such that there is no interaction with imaginary copies of particles) and make sure that the lattice discretization $\delta_\mathrm{s}$ is  small compared to the healing length in the center of the trap, $\xi_\mathrm{h}\approx\hbar/\sqrt{2m\mu}$, and large in comparison to the scattering length $a$ \cite{discretization}. Further details on the simulation parameters are given in \App{CLangevin}. 

In contrast to mean-field approaches, our full quantum simulations require a renormalization of the contact coupling $g$. In the CL framework, $g$ and 
$C_\text{dd}$ are input parameters, whereas the experimentally measurable quantity is the scattering length $a$~\cite{Pethick:2008, Pitaevskii:2016}.
To first order, $g$ and $a$ are related by $g=4\pi\hbar^2a/m$, but this relation acquires corrections of order $a/\delta_\mathrm{s}$ with $\delta_\mathrm{s}$ the spatial lattice spacing. This error cannot be made arbitrarily small by increasing $\delta_\mathrm{s}$ since $\delta_\mathrm{s}$ must remain sufficiently small compared to $\xi_{\rm h}$ in order to resolve the long-range physics properly, as mentioned above.
Beyond leading order, the relation between the bare interaction potential in Fourier space, $V_\mathbf{k}=g+C_\mathrm{dd}\left(k_z^2/|\mathbf{k}|^2-1/3\right)$, and $a$ is obtained from the Born series for the scattering amplitude, see \App{Born_series}.

The dispersion relation $\omega(\mathbf{k})$ of a Bose gas is in principle obtained from the dynamic structure factor (DSF).  As it includes dynamical properties, computing the DSF from imaginary-time simulations requires an analytic continuation~\cite{boninsegni:1996,vitali:2010}, which, as an ill-posed problem, is challenging. Assuming a single excitation branch, this difficulty can be avoided by various methods, see e.g.~\cite{Heinen2024a}, and in particular by relating the dispersion to the static structure factor (SSF),
\begin{align}
    \label{eq:SSF}
    S(\mathbf{k})
    =1+\frac{1}{N}
    \int_{\mathbf{r}_1,\mathbf{r}_2}\,
    \e^{\mathrm{i}\mathbf{k}\cdot(\mathbf{r}_1-\mathbf{r}_2)}\langle 
    \psi(\mathbf{r}_1)^\dagger \psi(\mathbf{r}_2)^\dagger
    \psi(\mathbf{r}_1) \psi(\mathbf{r}_2)\rangle\,.
\end{align}
The non-zero-temperature Feynman relation connects $S(\mathbf{k})$ to $\omega(\mathbf{k})$ via \cite{Pitaevskii:2016,Klawunn:2011}
\begin{align} 
\label{eq:Feynman_relation}
    S(\mathbf{k})=\frac{\hbar\,\mathbf{k}^2}{2m\,\omega(\mathbf{k})}\,\coth\left(\frac{\hbar\omega(\mathbf{k})}{2k_\mathrm{B}T}\right)\,,
\end{align}
see also Apps. \ref{app:feynmanrel} and \ref{app:Omrotdef}. As a simple four-point function, $S(\mathbf{k})$ can be straightforwardly evaluated in CL simulations, and one can thus obtain $\omega(\mathbf{k})$ with a standard root finding algorithm. 

\section{Results}
\begin{figure}[t]
	\centering
	\includegraphics[width=\columnwidth]{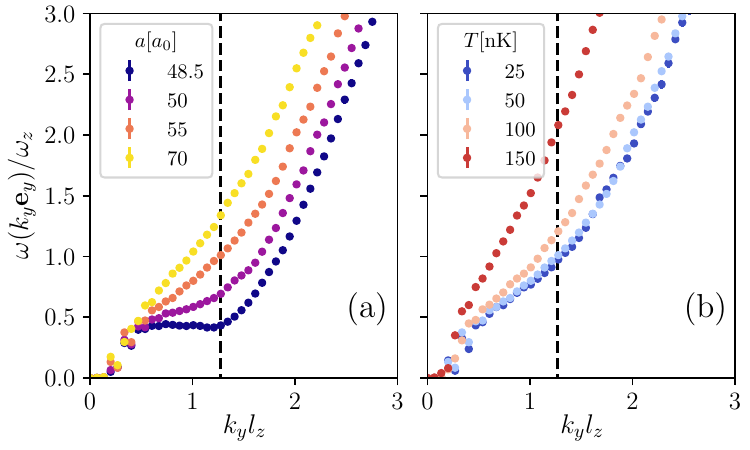}
    \caption{Dispersion relation $\omega(k_y\mathbf{e}_y)$ obtained from the SSF \eq{SSF} via the Feynman relation \eq{Feynman_relation}, for $N=2.4\times 10^4$ atoms, (a) for fixed $T=50\,\text{nK}$ and different scattering lengths $a$ as in \Fig{Sk}; (b) at fixed $a=55\,a_0$ for different $T$. The values for $a$ and $T$ are given in the respective legends.}
    \label{fig:omegak}
\end{figure}

While we can access $\omega(\mathbf{k})$ for arbitrary $\mathbf{k}$ in our CL simulations, we restrict further discussion to momenta along the weakly trapped $y$-direction, since this is the relevant axis for roton softening~\cite{Chomaz:2018,Petter:2019}. \Fig{Sk} show the SSF $S(k_y\mathbf{e}_y)$ extracted from the CL simulations for four values of $a$ and with $k_y$ given in units of the harmonic oscillator length in the direction of the dipole orientation $l_z=\sqrt{\hbar/m\omega_z}$ . 
The corresponding dispersion relations $\omega(k_y\mathbf{e}_y)$ obtained from \Eq{Feynman_relation} are presented in \Fig{omegak}(a). At low momenta, $0\leq k_yl_z\lesssim 0.5$, the phonon branches are approximately degenerate but difficult to resolve as $2\pi/k_y$ is on the order of system size, meaning that $k_y$ is not a good quantum number. At larger $k_y$, the branches smoothen and separate, exhibiting an overall softening with decreasing $a$. For $a \lesssim 50\,a_0$, a roton minimum appears, as reported in \cite{Petter:2019}.  

Our CL simulations also allow us to study the dependence of the dispersion on the gas parameters, in particular the temperature $T$, which is not easily accessible via standard mean-field calculations. Fig.~\fig{omegak}(b) shows results at fixed $a=55\,a_0$ and various $T$. For the chosen interaction parameter, increasing $T$ leads to larger $\omega$ for all $k$ beyond the phonon regime. This is apparently in contrast to the predictions of \cite{Sanchez2023heating}, but in agreement with the trend found in \cite{bisset2012finite}. 
The discrepancy may originate from the different constraints used in these approaches: while \cite{Sanchez2023heating} fixes the number of \textit{condensed} atoms, both \cite{bisset2012finite} and our calculations keep the \textit{total} atom number constant. Lowering the temperature below $50\,\text{nK}$ results only in a slight decrease of $\omega(\mathbf{k})$, indicating that thermal effects become almost negligible in the deeply degenerate regime. We emphasize that these results are obtained at a single scattering length away from the roton instability; exploring the temperature dependence in different interaction regimes, and in particular closer to the softening point, warrants further investigations.

\begin{figure}[t]
	\includegraphics[width=1.0\columnwidth]{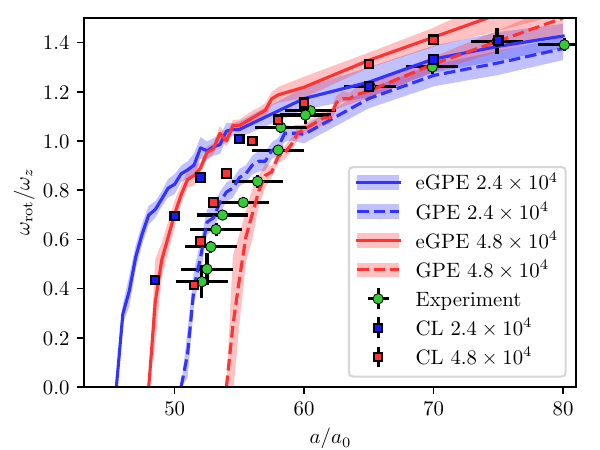}
	\caption{Roton gap $\omega_{\rm rot}$ as a function of the scattering length $a$ for fixed $T=50\,\text{nK}$ from ab initio Complex Langevin simulations (squares), compared against GPE (dashed) and eGPE (solid) simulations. Each set of simulations is performed at two particle numbers, $N=2.4\times10^4$ (blue) and $N=4.8\times10^4$ (red). Uncertainties in (e)GPE simulations are from the statistical spread of the dynamic structure factor peaks due to thermal noise. Green circles represent the experimental data from \cite{Petter:2019}, with mean condensed number $N_{\mathrm{BEC}}=2.4\times10^4$ and an average condensed fraction of approximately $50\%$ ~\cite{FFdiscussion}.
	}
	\label{fig:omrot}
\end{figure}

Following Ref.~\cite{Petter:2019}, we set the roton momentum to  $k_\text{rot}\equiv 1.27\, l_z^{-1}$, which is the momentum at which the dispersion touches zero, i.e.~the roton at the mean-field instability. We thereby define the roton gap at any scattering length as $\omega_\text{rot}\equiv \omega(k_\text{rot}\mathbf{e}_y)$, also when there is no local minimum yet. The roton gap extracted from the CL simulations, including those shown in \Fig{omegak}(a), is displayed in \Fig{omrot}, which demonstrates the softening of the roton mode with decreasing $a$.

As $a$ is lowered, an instability is ultimately crossed, identified by an unphysical negative peak in the static structure factor $S(\mathbf{k})$, reflecting imaginary values of the dispersion at the roton momentum. We can thus obtain an estimate for the scattering length at the instability, $a_\text{crit}$, by means of a lattice search in $a$ and $\mu$. We have $a>a_\text{crit}$ if there is a $\mu$ such that we reach the target particle number $N$ and $S$ remains positive, otherwise $a<a_\text{crit}$. We find $47.5\,a_0\leq a_\text{crit}\leq 48.5\,a_0$, resp. $51.0\,a_0\leq a_\text{crit}\leq 51.5\,a_0$, for $N=2.4\times 10^{4}$, resp. $N=4.8\times 10^{4}$.  

We compare the CL results against mean field simulations based on the (e)GPE,
\begin{align} 
\label{eq:EGPE}
  \mathrm{i}\hbar\frac{\partial \Psi}{\partial t}
  =\frac{\delta}{\delta \Psi^*}
  \int\mathrm{d}^3{r}\,\left\{ 
  \mathcal{H}\left[\Psi,\Psi^*\right]
  +\frac{2}{5}\gamma_{\text{QF}}|\Psi|^5\right\}\,,
\end{align}
by including (or not), in \eqref{eq:EGPE}, the LHY correction proportional to $\gamma_{\text{QF}}=128\sqrt{\pi}\hbar^2a^{5/2}\mathrm{Re}\left[\mathcal{Q}_5(\varepsilon_{\mathrm{dd}})\right]/3m$, with $\mathcal{Q}_5(x)=\int_0^1\mathrm{d}u\;(1-x+3u^2x)^{5/2}$~\cite{Schuetzhold:2006,Lima:2011,Lima:2012,Waechtler:2016,Chomaz:2016,Chomaz:2018,Petter:2019}, see \App{StructureFactor} for details. 
We use identical numerical grids as in our CL simulation. In doing so, we do not resort to Bogoliubov-de Gennes simulations as presented in \cite{Petter:2019}, which, on our large grids, are resource-intensive, and do not capture thermal effects, but employ real-time simulations close to equilibrium for probing the DSF. See, however, \App{RotonBog} for a brief description of the Bogoliubov-de Gennes approach. We employ a Truncated-Wigner-type approach by first calculating the ground state of the (e)GPE through imaginary-time propagation, then populating the system with stochastic noise at $T=50\,\text{nK}$, and allowing it to relax via real-time evolution. 
The DSF is then evaluated as
\begin{equation}
    S(\omega,\textbf{k})
    =\int\frac{\id t}{2\pi}\e^{\mathrm{i}\omega t}
    \langle\delta \tilde{n}(\textbf{k},t) 
    \delta \tilde{n}(\textbf{k},0)\rangle\;,
\end{equation}
where $\delta \tilde{n}(\textbf{k},t)$ are the momentum-space fluctuations of the density calculated with the (e)GPE, and $\langle...\rangle$ represents an average over many instances of the thermal noise. 
From the position of the peaks in $S(\omega,\textbf{k})$ for each $k\mathbf{e}_y$, including the roton momentum, we extract the shape of the dispersion and the roton gap, respectively, i.e., $\omega_\text{rot} = \text{max}[S(\omega,k_\mathrm{rot}\textbf{e}_y)]$. The results are depicted in \Fig{omrot} alongside the CL data. \Fig{GPEvsBog} in \App{StructureFactor} shows a comparison to computations based on Bogoliubov theory, with and without the LHY correction that were reported in \cite{Petter:2019}, which turn out to agree closely with our results, reaffirming the validity of our approach.

\Fig{omrot} displays simulation results for both $N=2.4\times10^4$ and $N=4.8\times10^4$, see discussion below. For both particle numbers, we observe a quantitative difference between the GPE and eGPE roton gaps, with the discrepancy becoming larger closer to the instability. At large $a$, far from any instability, the CL and eGPE results agree, meanwhile the GPE prediction consistently underestimates the roton gap. At lower $a$, however, the CL result deviates from the eGPE prediction and instead implies the roton gap to lie between the GPE and eGPE values. 
 
The LHY correction $\frac{2}{5}\gamma_{\text{QF}}|\Psi|^5$ entering the eGPE~\eq{EGPE} is derived within a local density approximation (LDA) and assuming a uniform particle density~\cite{Waechtler:2016,Lima:2012}. This is approximately justified since the LHY correction is dominated by the contribution of large-$k$ modes, which are barely sensitive to the trap inhomogeneity. Nevertheless, quantitative deviations can arise. A striking example of such inaccuracies is that the LDA yields an imaginary contribution to the LHY correction for $a<a_{\mathrm{dd}}$. This arises from the instability of the uniformly trapped BEC at $a=a_{\mathrm{dd}}$ and encompasses the contribution of the small-$k$ modes. The confined BEC is stabilized at the mean-field level thanks to the trapping potential, making this imaginary contribution unphysical~\cite{Baranov:2008}.  The imaginary part is typically small and therefore discarded in practical calculations, but still reaches a few percent of the real part at $a\approx 55\,a_0$, where CL and eGPE predictions are seen to start deviating from each other in \Fig{omrot}. The CL method could also be well suited to studying potential new functional forms of the quantum fluctuations, similar to a recent study~\cite{Bombin2024Quantum} using PIMC, and now extending the study to experimentally realistic particle numbers.

We finally compare our simulations to the experimental results of Ref.~\cite{Petter:2019}, also displayed in \Fig{omrot} (green circles). An average condensed particle number of $2.4\times 10^4$ was measured, while the average total particle number was twice as large, at $\sim4.8\times 10^4$~\cite{FFdiscussion}.
We therefore consider both values in our simulations: 
$N=2.4\times 10^4$, as in the theory of Ref.~\cite{Petter:2019} accounting only for the estimated condensed part, and  $N=4.8\times 10^4$, accounting for the full gas as should be captured in CL. We note that for the temperature estimated in experiment and used for our CL calculations (50\,nK), the estimated thermal fraction from theory disagrees with that extracted from experiments, and only sums up to a few percent, also potentially justifying the use of a larger atom number in (e)GPE simulations. 
Near the instability, up to approximately $a=60\,a_0$, the experimental and the CL data for $N=4.8\times 10^4$ (red squares) agree remarkably well, besides the slight bending of the experimental curve inwards around $a\approx57\,a_0$. Meanwhile, at higher scattering lengths the data diverges slightly, and the experiment appears to agree better with the CL evaluated with $N=2.4\times 10^4$ (blue squares). It should be noted that in the experiment, the atom number, temperature, and condensed fraction show significant variations from one parameter set ($a$ value) to the next~\cite{FFdiscussion} not accounted for neither by the experimental errorbars nor in the various theoretical computations. Therefore, comparing individual data points carries an intrinsic uncertainty. Additional experimental uncertainties, e.g., a systematic error on the scattering length values, a potential underestimation of the temperature (see comment on the thermal fraction above), etc., might also contribute to some of the discrepancies between our computations and the experiment. This comparison warrants further investigations both from theory and experiment, which are beyond the scope of the present work.

\section{Conclusions}
We have implemented an ab initio quantum Monte-Carlo simulation of a dipolar Bose gas near the roton instability using the CL algorithm, enabling access to experimentally relevant regimes. The CL simulation at nonzero temperature predicts a roton gap in agreement with eGPE theory for large $a$, while close to the instability the CL result lies in between GPE and eGPE values.  Comparison with the experiment of Ref.~\cite{Petter:2019} shows quantitative agreeement with the CL predictions when the total -- i.e.~condensed plus thermal -- atom number is taken into account, especially near the instability where quantum fluctuations are crucial, but the eGPE fails to capture the quantitative behavior. Our results highlight the need for an improved inclusion of beyond-mean-field quantum corrections to the GPE, especially for $\varepsilon_\mathrm{dd}>1$.

The numerically exact CL approach to dipolar Bose gases promises a wide range of future applications, as those gases are, in certain regimes, highly sensitive to quantum and thermal fluctuations beyond the established approximate descriptions. First, it provides a route to better understand the range of applicability of the approximate descriptions such as the eGPE, including the form of the LHY correction, the effect of different trapping geometries, and the role of temperature. Second, it is a powerful method for studying the interplay of interactions and thermal phase transitions \cite{bisset2011finite,bisset2012finite}. A promising application for dipolar systems is the physics beyond the roton instability, where supersolids and quantum droplets form~\cite{Tanzi:2019,Boettcher:2019,Chomaz:2019,Chomaz:2016,Schmitt:2016}. In this regime, quantum effects stabilize the gas against collapse \cite{Waechtler:2016}, such that a first-principle simulation is highly desirable.  While there have been such simulations within the PIMC framework \cite{Saito:2016,Cinti:2017,Bombin2025creating}, it would be interesting to employ CL simulations for exploring experimentally realistic particle numbers while including temperature, which has also recently been found to play an important role in the formation of the supersolid phase~\cite{Sanchez2023heating,LiangJun2025accessing}. A similar application of CL extends beyond dipolar gases to all systems with large $N$ where fluctuations play an essential role, including mixtures~\cite{Petrov:2015,Cabrera2018Quantum,Semeghini2018SelfBound}, spin-orbit coupled BECs~\cite{Li2017Stripe,Putra2020spatial,Chisholm2024pst}, or lattice systems~\cite{Gross2017Quantum}. 

\section{Acknowledgments}The authors thank Thomas Bland, Karthik Chandrashekara, Luca Falzoni, Francesca Ferlaino, Andreea Oros, Thomas Pohl, Elena Poli, Niklas Rasch, Alessio Recati, and Pramodh Senarath Yapa, for discussions and collaboration on related topics. 
This work was supported 
by the European Research Council (ERC) under the European Union’s Horizon Europe research and innovation program under grant number 101040688 (project 2DDip), by the Deutsche Forschungsgemeinschaft (DFG, German Research Foundation), through 
SFB 1225 ISOQUANT (Project-ID 273811115), 
grant GA677/10-1, 
and under Germany's Excellence Strategy -- EXC 2181/1 -- 390900948 (the Heidelberg STRUCTURES Excellence Cluster), 
as well as by the state of Baden-W{\"u}rttemberg through bwHPC, the data storage service SDS{@}hd supported by the Ministry of Science, Research and the Arts Baden-W{\"u}rttemberg (MWK), and the DFG through grants 
INST 35/1503-1 FUGG, INST 35/1597-1 FUGG, and INST 40/575-1 FUGG
(SDS, Helix, and JUSTUS 2 clusters). We gratefully acknowledge financial support from Heidelberg University towards the publication fee.

\clearpage

\appendix
\section{The Complex Langevin approach}  
\label{app:CLangevin}
%
\subsection{Complex Langevin equations}  
\label{app:CLangevinEqs}
The basic complex Langevin-type stochastic differential equations in a fictitious time $\vartheta$ take the form given in \Eq{cl_eq12} in the main text,
\begin{align}
    \label{eq:cl_eq12_app}
    \frac{\partial\psi}{\partial\vartheta}
    &=-\frac{1}{\hbar}\frac{\delta S}{\delta \bar{\psi}}+\eta(\vartheta)
    \,,\\
    \label{eq:cl_eq2}
    \frac{\partial\bar{\psi}}{\partial\vartheta}
    &=-\frac{1}{\hbar}\frac{\delta S}{\delta \psi}+\eta(\vartheta)^*
    \,,
\end{align}
through which the components $\varphi$ and $\chi$ of the fields $\psi=\varphi+\mathrm{i}\chi$, $\bar\psi=\varphi-\mathrm{i}\chi$ become complex. 
The complex noise fulfills 
\begin{align}
    \langle\eta(\vartheta)\rangle
    &=\langle\eta^*(\vartheta)\rangle = 0\,,
    \label{eq:noisecorr1}\\
    \langle\eta(\vartheta)\eta(\vartheta')\rangle
    &=\langle\eta^*(\vartheta)\eta^*(\vartheta')\rangle=0\,,
    \label{eq:noisecorr2}\\
  \langle\eta(\vartheta)\eta(\vartheta')^*\rangle
    &=2\delta(\vartheta-\vartheta')\,.
    \label{eq:noisecorr3}
\end{align}
By expressing $\eta(\vartheta)$ in terms of the uncorrelated noise variables $\eta_\varphi$ and $\eta_\chi$ of the fields as
\begin{align}
    \eta(\vartheta) = \eta_\varphi(\vartheta) + \mathrm{i}\eta_\chi(\vartheta)\,,
\end{align}
the latter noise variances can be rewritten as
\begin{align}
    \langle\eta(\vartheta)\eta(\vartheta')\rangle
    &=\langle\eta_\varphi(\vartheta)\eta_\varphi(\vartheta')
    -\eta_\chi(\vartheta)\eta_\chi(\vartheta')\rangle = 0\,,
    \nonumber\\
    \langle\eta(\vartheta)\eta(\vartheta')^*\rangle
    &=\langle\eta_\varphi(\vartheta)\eta_\varphi(\vartheta')
    +\eta_\chi(\vartheta)\eta_\chi(\vartheta')\rangle 
    = 2\delta(\vartheta-\vartheta')\,,
\end{align}
where we used that $\langle\eta_\varphi(\vartheta)\eta_\chi(\vartheta')\rangle=0$. 

The fact that we are dealing with only one complex noise variable $\eta$ (instead of two independent such variables $\eta$ and $\bar{\eta}$ as for the fields) is a consequence of our choice to take $\eta_\varphi$ and $\eta_\chi$ to be purely real, in contrast to the fields $\varphi$ and $\chi$. In principle, one can also take them to be complex \cite{Heinen:2022}, $\eta_\varphi = \eta_{\varphi,R}+\mathrm{i}\eta_{\varphi,I}$, $\eta_\chi = \eta_{\chi,R}+\mathrm{i}\eta_{\chi,I}$, and they must then fulfill
\begin{align}
\langle\eta_{\varphi,R}(\vartheta)\eta_{\varphi,R}(\vartheta')\rangle&=N_R\,\delta(\vartheta-\vartheta')\\
\langle\eta_{\varphi,I}(\vartheta)\eta_{\varphi,I}(\vartheta')\rangle&=N_I\,\delta(\vartheta-\vartheta')
\end{align}
and analogously for $\eta_{\chi,R}$ and $\eta_{\chi,I}$. The parameters $N_R$ and $N_I$ can be chosen arbitrarily under the constraint $N_R-N_I=1$, but the choice $N_R=1$, $N_I=0$, i.e. to take purely real noise variables $\eta_\varphi$ and $\eta_\chi$, is standard in the literature, as it is numerically most stable. Additionally, it allows to write the noise correlators in the simple form of \eq{noisecorr1}-\eq{noisecorr3}.

\subsection{Discretized Langevin equations for a\\ dipolar Bose gas \label{app:Langevin_eq}}
The discretized version of the action \eq{action} in the main text reads, with spatial and imaginary-time lattice spacing $\delta_\mathrm{s}$ and $\delta_\tau$, respectively:
\begin{align}
    \nonumber S_{\text{disc}}=&\ \delta_\mathrm{s}^3\sum_{i=0}^{N_\tau-1}\sum_\mathbf{j}\Bigg\{\hbar\bar{\psi}_{i+1,\mathbf{j}}\left(\psi_{i+1,\mathbf{j}}-\psi_{i,\mathbf{j}}\right)
    \\\nonumber&+\delta_\tau\Bigg[-\frac{\hbar^2}{2m}\bar{\psi}_{i+1,\mathbf{j}}\Delta^\text{lat}\psi_{i,\mathbf{j}}-\left(\mu-V^\text{ext}_\mathbf{j}\right)\,\bar{\psi}_{i+1,\mathbf{j}}\psi_{i,\mathbf{j}}
    \\&+\frac{g}{2}\left(\bar{\psi}_{i+1,\mathbf{j}}\psi_{i,\mathbf{j}}\right)^2+\frac{1}{2}U^\text{dip,lat}_{i,\mathbf{j}}\,\bar{\psi}_{i+1,\mathbf{j}}\psi_{i,\mathbf{j}}\Bigg]\Bigg\}\,,
    \label{eq:discaction}
\end{align}
where the index $i=0,\dots,N_\tau-1$ numbers the $N_\tau=\hbar\beta/\delta_\tau$ lattice points in imaginary time direction (due to the bosonic boundary condition, $i=N_\tau$ is equivalent to $i=0$) and the index vector $\mathbf{j}$ the spatial lattice points. Note that $\bar{\psi}$ must in general be evaluated one lattice point earlier than $\psi$, due to their different indices $i+1$ and $i$ in \eq{discaction} \cite{Heinen:2022}. We evaluate the lattice version of both the Laplacian, $\Delta^\text{lat}$, and the dipolar potential, $U^\text{dip,lat}$, in Fourier space. The discretized Laplacian reads
\begin{align}
    \Delta^\text{lat}\psi_{i,\mathbf{j}}
    =-\frac{1}{\mathcal{V}}\sum_{\mathbf{k}\mathbf{j}'}\exp\left[\mathrm{i}\mathbf{k}\cdot(\delta_\mathrm{s}\mathbf{j}-\delta_\mathrm{s}\mathbf{j}')\right]|\mathbf{k}|^2\psi_{i,\mathbf{j}'}\,,
\end{align}
where $\mathcal{V}=N_{x}N_{y}N_{z}$ is the lattice volume, the indices $\mathbf{j},\mathbf{j}'$ numerate the spatial lattice points as before and $\mathbf{k}$ runs through the lattice momenta, i.e., $\mathbf{k}=2\pi \delta_\mathrm{s}^{-1}\left(\ell_x/N_{x},\ell_y/N_{y},\ell_z/N_{z}\right)$, with $\ell_{x,y,z}$ running from $-N_{x,y,z}/2$ to $N_{x,y,z}/2$. The dipolar potential is evaluated as 
\begin{align}
    U^\text{dip,lat}_{i,\mathbf{j}}=&\frac{1}{\mathcal{V}}\sum_{\mathbf{k}\mathbf{j}'}\exp\left[\mathrm{i}\mathbf{k}\cdot(\delta_\mathrm{s}\mathbf{j}-\delta_\mathrm{s}\mathbf{j}')\right] \tilde{V}_\mathbf{k}^{\text{dip}}\,\bar{\psi}_{i+1,\mathbf{j}'}\psi_{i,\mathbf{j}'}\,,
\end{align}
where $\tilde{V}_\mathbf{k}^{\text{dip}}$ is the Fourier transform of a dipolar potential with long-distance cutoff at half the lattice extension, which we employ in order to avoid self-interactions of atoms with their imaginary copies:
\begin{align}
    \nonumber\tilde{V}_\mathbf{k}^{\text{dip}}=&\ \sum_{j_x=-N_x/2}^{N_x/2}\sum_{j_x=-N_y/2}^{N_y/2}\sum_{j_x=-N_z/2}^{N_z/2}\exp\left[\mathrm{i}\mathbf{k}\cdot(\delta_\mathrm{s}\mathbf{j}-\delta_\mathrm{s}\mathbf{j}')\right]\\&\times\frac{C_\mathrm{dd}}{4\pi}\frac{1-3(\delta_\mathrm{s}j_z)^2/|\delta_\mathrm{s}\mathbf{j}|^2}{|\delta_\mathrm{s}\mathbf{j}|^3}\,.
\end{align}
From the discretized action, one readily derives the Langevin equations for $\psi$ and $\bar\psi$ in the Langevin time $\vartheta$,
\begin{align}
    \nonumber\hbar\frac{\partial\psi_{i,\mathbf{j}}}{\partial\vartheta}=
    &\ \delta_\mathrm{s}^3\Bigg\{\hbar(\psi_{i-1,\mathbf{j}}-\psi_{i,\mathbf{j}})\\\nonumber
    &+\delta_\tau\Bigg[\frac{\hbar^2}{2m}\Delta^\text{lat}\psi_{i-1,\mathbf{j}}+\left(\mu-V^\text{ext}_\mathbf{j}\right)\psi_{i-1,\mathbf{j}}\\
    &-g\,\bar{\psi}_{i,\mathbf{j}}\psi_{i-1,\mathbf{j}}\psi_{i-1,\mathbf{j}}-U^\text{dip,lat}_{i-1,\mathbf{j}}\psi_{i-1,\mathbf{j}}\Bigg]\Bigg\}+\eta_{i,\mathbf{j}}\,,
    \\
    \nonumber\hbar\frac{\partial\bar{\psi}_{i,\mathbf{j}}}{\partial\vartheta}=
    &\ \delta_\mathrm{s}^3\Bigg\{\hbar(\bar{\psi}_{i+1,\mathbf{j}}-\bar{\psi}_{i,\mathbf{j}})\\\nonumber
    &+\delta_\tau\Bigg[\frac{\hbar^2}{2m}\Delta^\text{lat}\bar{\psi}_{i+1,\mathbf{j}}+\left(\mu-V^\text{ext}_\mathbf{j}\right)\bar{\psi}_{i+1,\mathbf{j}}\\
    &-g\,\bar{\psi}_{i+1,\mathbf{j}}\psi_{i,\mathbf{j}}\bar{\psi}_{i+1,\mathbf{j}}-U^\text{dip,lat}_{i,\mathbf{j}}\bar{\psi}_{i+1,\mathbf{j}}\Bigg]\Bigg\}+\eta_{i,\mathbf{j}}^*\,.
\end{align}
These equations are simulated by means of the first-order Euler-Maruyama method.

For the discretization, we use a $48\times 384\times 64\times N_\tau$ lattice ($N_\tau$ ranging from $14$ to $84$), which amounts to approximately twice the extension of the atom cloud in every direction (which is, even in the presence of a long-distance cutoff to the dipolar potential, necessary in order to exclude the effect of imaginary copies). For the ratio of imaginary-time and spatial discretization we choose $\delta_\tau=0.1\,m\delta_\mathrm{s}^2/\hbar$. This makes sure that even for the largest wave vectors $\mathbf{k}$ resolved by the lattice we still have $\delta_\tau\varepsilon(\mathbf{k})/2\hbar\ll1$ with the kinetic energy $\varepsilon(\mathbf{k})\equiv{\hbar^2\mathbf{k}^2}/{2m}$, which is necessary for the stability of the simulation, confer also the remark after \eq{int_disc_prop}. The spatial discretization $\delta_\mathrm{s}$ should be chosen such that $\delta_\mathrm{s}/\pi$ is  small compared to the healing length in the center of the trap, $\xi_\mathrm{h}\approx\hbar/\sqrt{2m\mu}$, and at the same time it should be large in comparison to the scattering length $a$. We choose $\delta_\mathrm{s}=0.12\,\mu m$ such that $a_\mathrm{dd}=1.1\,\delta_\mathrm{s}/12\pi$, which results in $\pi\xi_{\rm h}/\delta_\mathrm{s}\approx 5$ and $\delta_\mathrm{s}/\pi a\approx 10$.

The fields $\psi$ and $\bar{\psi}$ are initialized with the mean-field ground state. After a warm-up period of the Langevin process of $\Theta_0=0.1\cdot10^4\,\delta_\mathrm{s}^{-3}=0.0248\,a_\mathrm{dd}^{-3}$, we typically average observables over a total Langevin time $\Theta=1.4\cdot10^4\,\delta_\mathrm{s}^{-3}=0.3478\,a_\mathrm{dd}^{-3}$ \cite{Langevin_time}, as well as over $20$ independent runs, and over the imaginary time $\tau$ (exploiting the translation invariance in the latter). Errors are estimated from the variance over the statistically independent runs.
%
\section{Determination of the scattering length\label{app:Born_series}}
The s-wave scattering length $a$, which is the macroscopic measure of the strength of the contact interaction in a gas of bosonic atoms, is related only in leading order to the bare coupling $g$ as $a=mg/4\pi\hbar^2$. While this approximation is consistent in mean-field simulations, where the scattering amplitude is evaluated near zero momentum only, a full quantum simulation must take into account the higher-order corrections in terms of the bare interaction potential that enters the simulations.

Within scattering theory, one shows that the relation between the bare interaction potential, $V_\mathbf{k}=g+C_\mathrm{dd}\left(k_z^2/|\mathbf{k}|^2-1/3\right)$, and the scattering length $a$ can be expressed by the Born series,
\begin{align}
\label{eq:Born}
\nonumber\frac{4\pi\hbar^2 a}{m}=&\ V_0-\frac{m}{\hbar^2} \int \frac{\id^3k}{(2\pi)^3}\frac{V_{\mathbf{k}}V_{-\mathbf{k}}}{\mathbf{k}^2}\\
&+\frac{m^2}{\hbar^4}\int \frac{\id^3k}{(2\pi)^3}\int \frac{\id^3k'}{(2\pi)^3}\frac{V_{-\mathbf{k}}V_{\mathbf{k}'}V_{\mathbf{k}-\mathbf{k}'}}{\mathbf{k}^2\mathbf{k}'^2}+\mathcal{O}\left(V^4\right)\,,
\end{align}
where momentum integrals are understood to be restricted to the momenta resolved by the computational lattice, and the limit $V_0$ has to be combined with an angular average, which amounts to $V_0=g$.
The $\mathcal{O}(V^2)$ contribution can straightforwardly be performed analytically for a spherically symmetric momentum cutoff, i.e. $|\mathbf{k}|<\Lambda$, while the $\mathcal{O}(V^3)$ integral becomes difficult to treat analytically. Furthermore, a cubic computational lattice does actually not implement a spherical momentum cutoff but rather a cutoff in the single components of the momentum vector. It is, however, straightforward to evaluate \Eq{Born} numerically.

The Born series \eq{Born} provides the correct relationship between s-wave scattering length $a$ and the interaction potential $V_\mathbf{k}$ for a lattice theory with spatial lattice discretization and hence a finite momentum cutoff. However, in our numerical simulations, we need to also discretize the imaginary time $\tau$. While it would be possible to make the imaginary time lattice spacing $\delta_\tau$ sufficiently small to reach convergence such that the Born series is fulfilled to a sufficient precision, we found this to be numerically very expensive as the convergence is comparatively slow. Hence, a better approach is to look for a modified version of the Born series that provides a valid expression for the scattering length also in the case of a discretized imaginary time. 

In field theory, the scattering length $a$ is given by the negative of the vacuum scattering amplitude $f$ of two particles in the limit of zero momentum, i.e. $a=-\lim_{\mathbf{k},\mathbf{k}'\to 0}f(\mathbf{k},\mathbf{k}')$. The requirement that the scattering happens in vacuum amounts to $T=\mu=0$ in field theory. The scattering amplitude is related to the $T$-matrix as $f(\mathbf{k},\mathbf{k}')=-m/(4\pi\hbar^2)T(\mathbf{k},\mathbf{k}')$. The latter in turn is related via the Lehmann–Symanzik–Zimmermann reduction theorem to correlation functions of the fields with truncated external propagators. The scattering length $a$ can thus be represented by Feynmann diagrams as:
\begin{align}
a=\begin{tikzpicture}[baseline={([yshift=-0.75ex]current bounding box.center)},vertex/.style={anchor=base,
	circle,fill=black!25,minimum size=18pt,inner sep=2pt}]
\draw[black,fill=black] (0,0) circle (0.5ex);
\draw[black] (0,0) -- (0.2,0.2);
\draw[black] (0,0) -- (-0.2,0.2);
\draw[black] (0,0) -- (0.2,-0.2);
\draw[black] (0,0) -- (-0.2,-0.2);
\end{tikzpicture}
\quad+\quad
\begin{tikzpicture}[baseline={([yshift=-0.75ex]current bounding box.center)},vertex/.style={anchor=base,
	circle,fill=black!25,minimum size=18pt,inner sep=2pt}]
\draw[black,fill=black] (0,0) circle (0.5ex);
\draw[black,fill=black] (1,0) circle (0.5ex);
\draw[black] (0,0) -- (-0.2,-0.2);
\draw[black] (0,0) -- (-0.2,0.2);
\draw[black] (1,0) -- (1.2,-0.2);
\draw[black] (1,0) -- (1.2,0.2);
\draw[black](0,0) .. controls (0.3,0.3) and (0.7,0.3) .. (1,0);
\draw[black](0,0) .. controls (0.3,-0.3) and (0.7,-0.3) .. (1,0);
\end{tikzpicture}
\quad+ \dots\,,
\end{align}
where the vertices denote the bare interactions and full lines between vertices represent (free) propagators.
The leading-order expression for the scattering length $a^\text{(LO)}$ is thus given by 
\begin{align}
   \frac{4\pi\hbar^2 a^\text{(LO)}}{m}=V_0\,.
\end{align}
For the next-to-leading order $a^\text{(NLO)}$ we have
\begin{align}
\label{eq:NLO}
   \frac{4\pi\hbar^2 a^\text{(NLO)}}{m}=-\int \frac{\hbar\id\omega}{2\pi}\int \frac{\id^3k}{(2\pi)^3}\,V_{\mathbf{k}}V_{-\mathbf{k}}\,G(\omega,\mathbf{k})G(-\omega,-\mathbf{k})\,.
\end{align}
For a continuum imaginary time, $\delta_\tau\to 0$, the propagator $G(\omega,\mathbf{k})$ reads
\begin{align}
   G(\omega,\mathbf{k})=\frac{1}{\mathrm{i}\hbar\omega+\varepsilon(\mathbf{k})}\,, 
\end{align}
with $\varepsilon(\mathbf{k})\equiv{\hbar^2\mathbf{k}^2}/{2m}$ and the integral over $\omega$ runs from $-\infty$ to $+\infty$. For a finite $\delta_\tau$, we obtain a discretized action and thus the propagator reads
\begin{align}
    G^\text{disc}(\omega,\mathbf{k})=\frac{1}{\hbar \delta_\tau^{-1}(\e^{\mathrm{i}\delta_\tau\omega}-1)+\varepsilon(\mathbf{k})}\,,
\end{align}
and the integral over $\omega$ runs from $-\pi/\delta_\tau$ to $+\pi/\delta_\tau$. 

Let us evaluate the frequency integral in \eq{NLO} for the discretized case:
\begin{widetext}
\begin{align}
      \label{eq:int_disc_prop}
      \nonumber
      \int \limits_{-\pi/\delta_\tau}^{+\pi/\delta_\tau}\frac{\hbar\id\omega}{2\pi}\,G(\omega,\mathbf{k})^\text{disc}G(-\omega,-\mathbf{k})^\text{disc}
      &=\int \limits_{-\pi/\delta_\tau}^{+\pi/\delta_\tau}\frac{\hbar\id\omega}{2\pi}\,\frac{1}{\hbar \delta_\tau^{-1}(\e^{\mathrm{i}\delta_\tau \omega}-1)+\varepsilon(\mathbf{k})}
      \frac{1}{\hbar \delta_\tau^{-1}(\e^{-\mathrm{i}\delta_\tau \omega}-1)+\varepsilon(\mathbf{k})}
      \\\nonumber
      &= \frac{1}{4\pi\varepsilon(\mathbf{k})}\int \limits_{-\pi}^{+\pi}\frac{\id x}{\left[1-{\hbar \delta_\tau^{-1}\varepsilon(\mathbf{k})^{-1}}\right]\cos(x)+{\hbar \delta_\tau^{-1}\varepsilon(\mathbf{k})^{-1}}-1+\delta_\tau\varepsilon(\mathbf{k})/2\hbar}
      \\
      &=\ \frac{m}{\hbar^2\mathbf{k}^2}\frac{1}{1-\delta_\tau\varepsilon(\mathbf{k})/2\hbar}\,.
\end{align}
Here, we have implicitly assumed that $\delta_\tau\varepsilon(\mathbf{k})/2\hbar<1$. Inserting this result into \eq{NLO}, we obtain for the next-to-leading contribution to the scattering length for discretized imaginary time, 
\begin{align}
    \label{eq:NLO_disc}
   \frac{4\pi\hbar^2 a^\text{(NLO),disc}}{m}=-\frac{m}{\hbar^2}\int \frac{\id^3k}{(2\pi)^3}\,\frac{V_{\mathbf{k}}V_{-\mathbf{k}}}{\mathbf{k}^2}\frac{1}{1-\delta_\tau\varepsilon(\mathbf{k})/2\hbar}\,.
\end{align}
For fixed momentum cutoff and $\delta_\tau\to 0$, this expression goes over into the NLO contribution in \eq{Born}, as it has to. Similarly, one finds for the NNLO contribution,
\begin{align}
    \label{eq:NNLO_disc}
    \frac{4\pi\hbar^2 a^\text{(NNLO),disc}}{m}=&\ \frac{m^2}{\hbar^4}\int \frac{\id^3k}{(2\pi)^3}\int \frac{\id^3k'}{(2\pi)^3}\frac{V_{-\mathbf{k}}V_{\mathbf{k}'}V_{\mathbf{k}-\mathbf{k}'}}{\mathbf{k}^2\mathbf{k}'^2}
    \frac{1}{1-\delta_\tau\varepsilon(\mathbf{k})/2\hbar}\frac{1}{1-\delta_\tau\varepsilon(\mathbf{k}')/2\hbar}\,.
\end{align}
\end{widetext}

For obtaining the results presented in the main text, we evaluate the modified Born corrections defined by Eqs.~\eq{NLO_disc} and \eq{NNLO_disc} numerically by replacing the momentum integrals by momentum sums in the standard way, discretizing on a $128^3$ ($\mathcal{O}(V^2)$ contribution) or $64^3$ ($\mathcal{O}(V^3)$ contribution) lattice. While the $\mathcal{O}(V^2)$ contribution results of the order of $\sim 10\%$ and thus is still quite significant for the parameters chosen here ($\pi\xi_{\rm h}/\delta_\mathrm{s}\approx 5$), the $\mathcal{O}(V^3)$ contribution becomes already almost negligible, amounting to shifts smaller than $1\%$.

\section{Range of validity of the Feynman relation\label{app:feynmanrel}}
In the CL calculations, we extract the dispersion relation from the \textit{static} structure factor (SSF) $S(\mathbf{k})$ via the non-zero-temperature Feynman relation, \eq{Feynman_relation} in the main text. This differs from the approach taken in both \cite{Petter:2019} and the benchmark (e)GPE simulations provided in this paper, as well as in other mean-field-based works, where  the peaks in the \textit{dynamic} structure factor (DSF) $S(\omega,\mathbf{k})$ are taken as a measure for the dispersion relation. Here, we want to briefly comment on the range of validity of our SSF approach. 

We note that the SSF is defined, in terms of the DSF, as 
\begin{align}
    S(\mathbf{k})
    \equiv N^{-1}\int^\infty_\infty \mathrm{d}\omega\, 
    S(\omega,\mathbf{k})
    \,.
\end{align}
As the starting point of our discussion, we take the exact f-sum rule~\cite{Pitaevskii:2016},
\begin{align}
\int\limits^\infty_{-\infty} \id\omega\, \hbar\omega\, S(\omega, \mathbf{k})= N \frac{\hbar^2k^2}{2m}\,.
\end{align}
This relation is derived on the basis of a few very general assumptions. In particular, the momentum of the particles is taken to enter the Hamiltonian only via the kinetic energy, and the system is assumed to be invariant under parity and time reversal. This is not fulfilled, e.g., for spin-orbit coupled or rotating BECs but is the case for the system studied here. In particular, no assumptions about the temperature or the interaction strength have to be made for the validity of the f-sum rule. 

Let us first consider the case of $T=0$. Using that $S(\omega,\mathbf{k})\equiv0$ for $\omega<0$ at $T=0$, as well as the definition of the SSF, we obtain:
\begin{align}
\label{eq:zeroT}
\frac{\int\limits^\infty_0 \id\omega\, \hbar\omega\, S(\omega, \mathbf{k})}{\int\limits^\infty_0 \id\omega\, S(\omega, \mathbf{k})}=\frac{\hbar^2k^2}{2m S(\mathbf{k})}\,.
\end{align}
This means that dividing the free particle dispersion by $S(\mathbf{k})$ (right hand side of \eq{zeroT}) gives just the mean value for $\omega$ at a certain momentum $\mathbf{k}$ weighted by the DSF $S(\omega, \mathbf{k})$ (left hand side of \eq{zeroT}), i.e., the ``center of mass'' of $S$. If $S(\omega, \mathbf{k})\propto \delta(\omega-\omega_\mathbf{k})$, we get $\hbar^2k^2/2m S(\mathbf{k})=\hbar\omega_\mathbf{k}$. The same is true also for a Gaussian, $S(\omega, \mathbf{k})\propto \exp(-(\omega-\omega_\mathbf{k})^2/2\sigma^2)$, and basically any peak shape symmetrically centered around $\omega_\mathbf{k}$.

For $T>0$ the relationship becomes slightly more complicated. Using $S(\omega,-\mathbf{k})=S(\omega,\mathbf{k})$ and $S(\omega,\mathbf{k})=\e^{\beta\hbar\omega}S(-\omega,-\mathbf{k})$, we obtain the relation
\begin{align}
\label{eq:finiteT}
\frac{\int\limits^\infty_0 \id\omega\, \hbar\omega\, 
(1-\e^{-\beta\hbar\omega})\,
S(\omega, \mathbf{k})}{\int\limits^\infty_0 \id\omega\,
(1+\e^{-\beta\hbar\omega})\, 
S(\omega, \mathbf{k})}=\frac{\hbar^2 k^2}{2m S(\mathbf{k})}
\end{align}
as the thermal generalization of \eq{zeroT}. For a delta function, $S(\omega, \mathbf{k})\propto \delta(\omega-\omega_\mathbf{k})$, this now gives the non-zero-temperature Feynman relation:
\begin{align}
\label{Feynman}
\hbar\omega_\mathbf{k}\tanh(\beta\hbar\omega_\mathbf{k}/2)=\frac{\hbar^2k^2}{2m S(\mathbf{k})}\,,
\end{align}
which is thus valid for arbitrary temperatures and interaction strength as long as the delta-function approximation remains valid.  Unfortunately, if $S(\omega,\mathbf{k})$ does not take the form of a delta distribution but has, e.g., the shape of a Gaussian with non-zero width, the Feynman relation \eqref{Feynman} does not follow exactly any more from the exact formula \eqref{eq:finiteT} at nonzero temperatures.

For $T\to 0$ we recover again the previous result that the Feynman relation is exact even for a Gaussian. We can examine how large the deviation can be at most for nonzero temperatures by examining the large-temperature limit, i.e., for $T\to \infty$. To see this, we expand \eqref{eq:finiteT} to lowest order in $\beta$ and insert a Gaussian with width $\sigma$ and mean $\omega_\mathbf{k}$ for $S(\omega,\mathbf{k})$. Assuming $\sigma\ll\omega_\mathbf{k}$, we obtain for the relative deviation between the dispersion $\tilde{\omega}_\mathbf{k}$ calculated according to \eqref{Feynman} and the ``true'' dispersion $\omega_\mathbf{k}$:
\begin{align}
\label{eq:error}
\frac{\tilde{\omega}_\mathbf{k}-\omega_\mathbf{k}}{\omega_\mathbf{k}}\approx\frac{1}{2}\left(\frac{\sigma}{\omega_\mathbf{k}}\right)^2\,.
\end{align}
Due to the quadratic dependence on $\sigma/\omega_\mathbf{k}$, this error bound quickly becomes very small. E.g., for the system considered here, $\sigma/\omega_\mathbf{k}\approx 0.1$ and thus the error is below $1\%$, smaller than the statistical error and hence negligible.

We can thus conclude: At $T=0$, the extraction of $\omega_\mathbf{k}$ via the Feynman relation corresponds \textit{exactly} to a ``center of mass'' definition of $\omega_\mathbf{k}$, while at $T>0$ this extraction method leads to small deviations from this definition, which are typically negligible, however. Deviations between $\omega_\mathbf{k}$ defined as the ``center of mass'' of $S(\omega, \mathbf{k})$ and other definitions, e.g. a ``location of maximum'' definition, are discussed in \App{Omrotdef}.
\section{Structure factor within Gross-Pitaevskii theory \label{app:StructureFactor}}
\begin{figure}[t]
	\includegraphics[width=0.95\columnwidth]{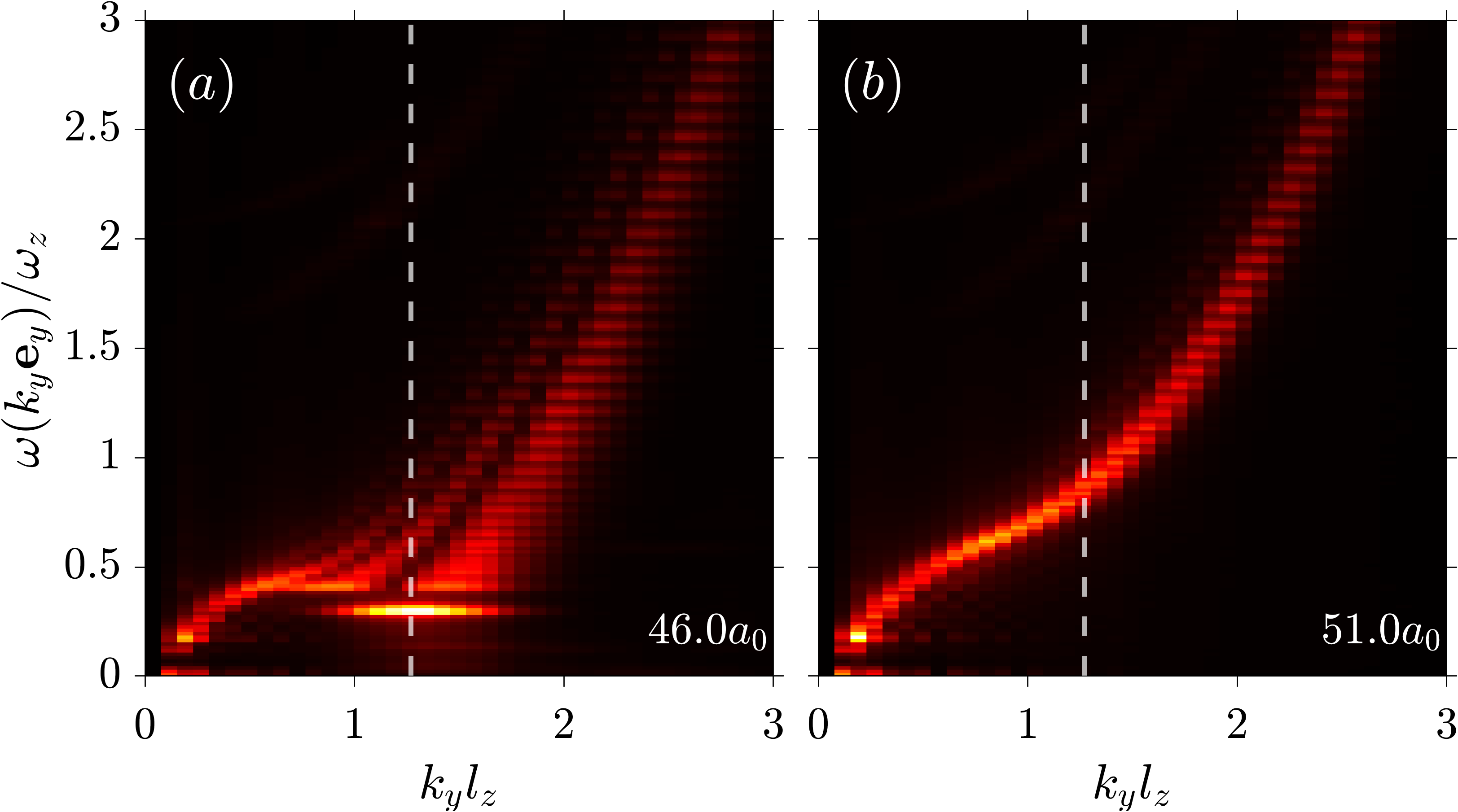}
	\caption{DSF calculated using eGPE theory for $N=2.4\times 10^{4}$, at two different scattering lengths displayed in the lower right corner of each panel. The roton momentum $k_\mathrm{rot}=1.27\,l_z^{-1}$ is marked with a vertical dashed line, and $\omega_\mathrm{rot}$ is determined by the peak signal along this line.
	}
	\label{fig:StaticStruc}
\end{figure}

In the benchmark (e)GPE simulations presented in the main text, we use the \textit{dynamic} structure factor (DSF) $S(\omega,\mathbf{k})$ to extract the dispersion relation. Here we detail the principle of this scheme. Generally speaking, the DSF at non-zero temperatures reads \cite{Pitaevskii:2016,Petter:2019},
\begin{equation}
    S(\omega,\textbf{k})=\mathcal{Z}^{-1}\sum_{ij}\e^{-\beta E_j}|\langle i|\delta \hat{n}_\textbf{k}|j\rangle|^2\delta(\omega-\omega_{i}+\omega_j)\;,
\end{equation}
where $\omega_i$ is the energy of the mode $i$, $\mathcal{Z}=\sum_{i}\e^{-\beta E_i}$ is the (canonical) partition function, and $\delta\hat{n}_\textbf{k}$ is the momentum-space density fluctuation operator with respect to the ground state $|0\rangle$,
\begin{equation}
    \delta\hat{n}_\textbf{k}=\int \frac{\id^3k}{(2\pi)^3}\;\e^{-\mathrm{i}\textbf{k}\cdot \textbf{r}}\left(\hat{\psi}^\dagger(\textbf{r})\hat{\psi}(\textbf{r})-\langle 0|\hat{\psi}^\dagger(\textbf{r})\hat{\psi}(\textbf{r})|0\rangle\right)\;.
\end{equation}
The DSF can be rewritten as, 
\begin{equation}
    S(\omega,\textbf{k})=\int\frac{\id t}{2\pi}\e^{\mathrm{i}\omega t}\langle\delta\hat{n}_\textbf{k}(t) \delta\hat{n}_\textbf{k}(0)\rangle_T\;,\label{eq:StrucFacExp}
\end{equation}
where the thermal expectation value is $\langle ...\rangle_T\equiv \sum_i \e^{-\beta E_i}\langle i|...|i\rangle$. 

In order to approximate the expectation value in \eq{StrucFacExp} within the semiclassical approximation of the (e)GPE, we calculate,
\begin{equation}
    S(\omega,\textbf{k})=\int\frac{\id t}{2\pi}\e^{\mathrm{i}\omega t}\delta \tilde{n}(\textbf{k},t) \delta \tilde{n}(\textbf{k},0)\;,\label{eq:SemiClassSF}
\end{equation}
where the density fluctuations are $\delta \tilde{n}(\textbf{k},t)=\tilde{n}(\textbf{k},t)-\tilde{n}_0(\textbf{k})$. Here, $\tilde{n}_0(\textbf{k})=|\tilde{\Psi}_0(\textbf{k})|^2$ is the the momentum density of the ground state of the (e)GPE with wavefunction $\Psi_0(\textbf{r})$ and $\tilde{n}(\textbf{k},t)=|\tilde{\Psi}(\textbf{k},t)|^2$ is time-dependent momentum-space density of the wavefunction $\Psi(\textbf{r}, t)$. The momentum dependence is obtained by spatial Fourier transform:
\begin{equation}
    \tilde{\Psi}(\textbf{k},t)=\int \mathrm{d}^3r\;\e^{-\mathrm{i}\textbf{k}\cdot\textbf{r}}\,\Psi(\textbf{r},t)\;.
\end{equation}

To get an estimate of the DSF, we employ a Truncated-Wigner-like approach where the wavefunction $\Psi(\textbf{r}, t)$ probes the time evolution of an initial condensate with thermal noise perturbation. We set
\begin{equation}
	\Psi(\textbf{r},t=0)=\Psi_0(\textbf{r})+\sideset{}{'}\sum_{nlm}\alpha_{nlm}\phi_{n}(x)\phi_{l}(y)\phi_{m}(z)\,,
\end{equation}
where the $\phi_i$ correspond to harmonic oscillator eigenstates, the complex amplitudes $\alpha_{nlm}$ are Gaussian random variables that obey $\langle |\alpha_{nlm}|^2\rangle=(\mathrm{e}^{(\epsilon_{nlm}-\mu)/k_\mathrm{B} T}-1)^{-1}$, where we use the trapping oscillator energies as $\epsilon_{nlm}=\hbar\omega_x(n+\frac{1}{2})+\hbar\omega_y(l+\frac{1}{2})+\hbar\omega_z(m+\frac{1}{2})$ and $n,l,m\in\mathbb{N}_0$. The sum $\sum'$ is restricted to states such that $\epsilon_{nlm}<2\,k_BT$. $\Psi(\textbf{r},t)$ is then evolved according to the (e)GPE for a total of $t=1\,\text{s}$.

The structure factor in \eq{SemiClassSF} is then averaged over 150 simulation runs. Each simulation run is prepared with independent instances of the initial thermal noise in order to approximate the expectation value. Each simulation run uses identical grid sizes as in the CL simulations and $T=50\,\text{nK}$.

\Fig{StaticStruc} shows examples of the DSF calculated for two different scattering lengths. Following Ref.~\cite{Petter:2019}, we extract the peak signal at $k_\mathrm{rot}=1.27\,l_z^{-1}$ to compare against the complex Langevin data in Fig. 3 in the main text. The increased signal in the DSF of Fig. 3 near $\omega_{\text{rot}}\approx 1.0\,\omega_z$ likely arises from strong signals due to the thermal population of the harmonic oscillator modes. This feature does not appear in every individual trial, and therefore depends on the initial random noise.

\begin{figure}
    \centering
    \includegraphics[width=0.9\columnwidth]{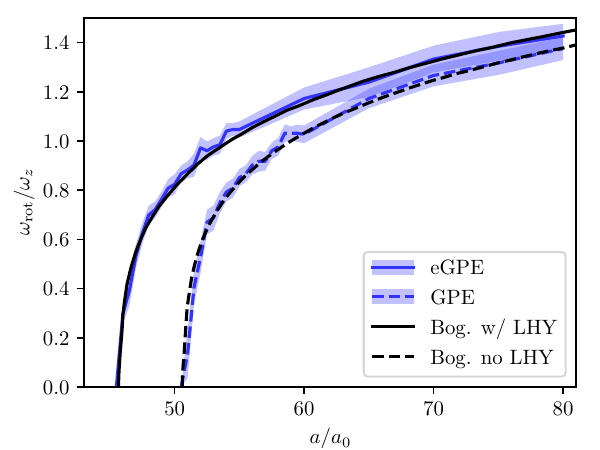}
    \caption{Roton mode softening similar to Fig. 3 in the main text, demonstrating that the dynamic evaluation of the structure factor \eq{SemiClassSF} matches the zero-temperature Bogoliubov theory from \cite{Petter:2019}, with and without LHY corrections.}
    \label{fig:GPEvsBog}
\end{figure}

For comparison, \Fig{GPEvsBog} shows the original results from \cite{Petter:2019} for $N=2.4\times 10^{4}$ based on a numerical solution of the Bogoliubov-de-Gennes equations alongside the results from our approach. They appear identical within errors, reaffirming the validity of our method. Also, this finding supports our hypothesis from the main text that $50$\,nK is indeed close enough to $T=0$ to compare these results on the same footing, although one must be cautious about how well the addition of thermal Truncated-Wigner to the initial condition captures the full thermal effects.

\section{Roton gap in Bogoliubov theory \label{app:RotonBog}}

For completeness, we show here how the roton gap can be extracted within Bogoliubov theory. For a system obeying the (e)GPE, the Bogoliubov-de Gennes (BdG) equations are obtained by considering linear field perturbations,
\begin{equation}
   \Psi(\textbf{r})=\left\{\Psi_0(\textbf{r})+\lambda\left[u(\textbf{r})\mathrm{e}^{-\mathrm{i}\omega t}-v^*(\textbf{r})\mathrm{e}^{\mathrm{i}\omega^* t}\right]\right\}\mathrm{e}^{-\mathrm{i}\mu t/\hbar}\,,
\end{equation}
in the form,
\begin{equation}
    \begin{pmatrix}
        H_0+C+X&-X\\
        X&-H_0-C-X
    \end{pmatrix}\begin{pmatrix}
        u\\
        v
    \end{pmatrix}=\hbar\omega\begin{pmatrix}
        u\\
        v
        \end{pmatrix}
    \,,
\end{equation}
where the operators are defined as follows,
\begin{align}
    H_0w(\textbf{r})=&\;-\frac{\hbar^2\nabla^2}{2m}+V_{\text{ext}}(\textbf{r})-\mu\\
    Cw(\textbf{r})=&\;\gamma_{\mathrm{QF}}|\Psi(\textbf{r})|^3w(\textbf{r})\nonumber\\&+\int\mathrm{d}\textbf{r}'V_{\text{int}}(\textbf{r}-\textbf{r}')|\Psi_0(\textbf{r}')|^2w(\textbf{r})\\
    X[w(\textbf{r})]=&\;\frac{3}{2}\gamma_{\mathrm{QF}}|\Psi_0(\textbf{r})|^3w(\textbf{r})\nonumber\\&\;+\Psi_0(\textbf{r})\int\mathrm{d}\textbf{r}'V_{\text{int}}(\textbf{r}-\textbf{r}')\Psi_0(\textbf{r}')w(\textbf{r}')
    \,,
\end{align}
and the interaction potential in (e)GPE theory reads,
\begin{equation}
    V_{\text{int}}(\textbf{r})=g\delta(\textbf{r})+\frac{C_\mathrm{dd}}{4\pi}\frac{1-3(z-z')^2/|\mathbf{r}-\mathbf{r}'|^2}{|\mathbf{r}-\mathbf{r}'|^3}
    \,.
\end{equation}
The dynamic structure factor (normalized to $N$) is,
\begin{equation}
    S(\omega,\textbf{k})=\frac{1}{N}\sum_i|\delta \tilde{n}(\textbf{k})|^2\delta(\omega-\omega_i)\label{eq:Skw_bare}
\end{equation}
where $i$ indexes all of the eigenenergies $\omega_i$ corresponding to solutions of the BdG equations, and $\tilde{n}(\textbf{k})$ is the Fourier-transform of the density perturbation,
\begin{equation}
    \delta n_i(\textbf{r})=\Psi_0(\textbf{r})[u_i(\textbf{r})-v_i(\textbf{r})]
    \,,
\end{equation}
with the BdG modes $u_i$ and $v_i$.

\begin{figure}
    \centering
    \includegraphics[width=0.95\columnwidth]{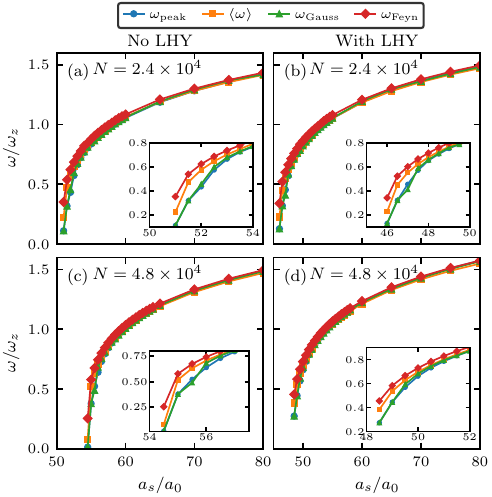}
    \caption{Roton mode softening using Bogoliubov theory. In the text, we describe the four approaches (i)-(iv) to calculate the roton gap using the dynamic structure factor.}
    \label{fig:BdGroton}
\end{figure}

In order to compare with the experiment, instead of using the bare structure factor in Eq.~\eqref{eq:Skw_bare}, we follow the same approach as in Ref.~\cite{Petter:2019} and slightly broaden the distribution by the Bragg pulse width $\tilde{\tau}$, approximating the DSG as,
\begin{equation}
    S(\omega, \textbf{k})\approx \frac{1}{N}\sum_i|\delta \tilde{n}(\textbf{k})|^2\tilde{\tau}\mathrm{sinc}^2(\tilde{\tau}(\omega-\omega_i)/2)\;.\label{eq:Skw_broad}
\end{equation}
%

\section{Comparison of different definitions for $\omega_\mathbf{k}$ \label{app:Omrotdef}}
In \App{feynmanrel}, we saw that an extraction of $\omega_\mathbf{k}$ from the static structure factor with the Feynman relation is equivalent to defining $\omega_\mathbf{k}$ as the center of mass of the dynamic structure factor $S(\omega, \mathbf{k})$. However, if one has direct access to $S(\omega, \mathbf{k})$, it is also possible to define $\omega_\mathbf{k}$ differently than as its center of mass. For example, another natural definition is to take the location of the maximum of $S(\omega, \mathbf{k})$ at fixed $\mathbf{k}$ as $\omega_\mathbf{k}$, or, especially in an experimental setting with noisy data, to fit a Gaussian or some other peak shape to $S(\omega, \mathbf{k})$ and define its center as $\omega_\mathbf{k}$ (as was done in Ref.~\cite{Petter:2019}). These definitions will yield the exact same value if the peak in $S(\omega, \mathbf{k})$ is symmetrically centered around $\omega_\mathbf{k}$. The latter is, however, not guaranteed, and for an asymmetric peak shape there will be deviations between these methods. In this section, we examine how large these deviations are in the considered scenario of the roton softening and thus whether the choice of definition can realistically impact the results.

Using the Bogoliubov method outlined in \App{RotonBog} (as it yields less noisy data compared to the Gross-Pitaevskii method described in \App{StructureFactor}, albeit at the cost of not capturing thermal effects), we take four independent approaches to calculating the roton gap shown in \Fig{GPEvsBog}. The first three correspond to the above-mentioned definitions; the fourth, which is based on \eq{finiteT}, is added in order to corroborate our estimate \eq{error} that the finite-$T$ deviations between a center-of-mass definition of $\omega_\text{rot}$ and its extraction via the Feynman relation are negligible. In each case, the roton momentum is fixed to $k_\mathrm{rot}=1.27\,l_z^{-1}$ as discussed elsewhere in the paper. The four approaches are defined as follows:

(i) First, we define the roton gap as $\omega_{\mathrm{peak}}$, the absolute peak of the structure factor $S(\omega, k_\mathrm{rot})$.

(ii) Second, we consider the ``center-of-mass'' via,
\begin{equation}
    \langle \omega\rangle = \frac{\int_{0}^{\omega_{\mathrm{max}}}\mathrm{d}\omega\:\omega\:S(\omega, k_\mathrm{rot})}{\int_{0}^{\omega_{\mathrm{max}}}\mathrm{d}\omega\:S(\omega, k_\mathrm{rot})}
\end{equation}
where $\omega_{\mathrm{max}}\gg \omega_z$ is the maximum BdG mode available via our numerical solver.

(iii) Third, similar to the experiment in Ref.~\cite{Petter:2019}, we fit a one-dimensional Gaussian to $S(\omega, k_\mathrm{rot})$. From this fit, we define the roton gap as the peak of the Gaussian $\omega_{\mathrm{Gauss}}$.

(iv) Finally we apply our approach using the finite-temperature Feynman relation, described in the main text and in \App{feynmanrel}. We calculate the structure factor in Eq.~\eqref{eq:Skw_broad}, and insert it into the finite-temperature expression given on the left-hand side of Eq.~\eqref{eq:finiteT}, using $T=50\,\mathrm{nK}$. We then implicitly solve for the roton gap by equating the left-hand sides of Eqs.~\eqref{eq:finiteT} and \eqref{Feynman}, which gives $\omega_{\mathrm{Feyn}}$.

The results of these alternatives, (i)-(iv), are shown in Fig.~\ref{fig:BdGroton} for each set of parameters studied in the main text. It becomes clear that, while the various methods of extraction of the roton gap yield slight variations, particularly within $\sim 1a_0$ of the instability (see insets), those differences are small compared to the variations across parameter ranges -- see e.g.~comparison between Fig.~\ref{fig:BdGroton}(a,b) and (c,d) -- as well as between GPE, eGPE, experiment, and CL -- see e.g.~comparison between Fig.~\ref{fig:BdGroton}(a,c) and (b,d). The largest variations between the curves occur close to the instability, where even the slightest differences in parameters can lead to strong variations. Meanwhile, the general trend tends to indicate that the instability does not shift by any appreciable amount when estimated by these various approaches. Furthermore, the relative proximity of $\omega_{\mathrm{Feyn}}$ to the three $T=0$ variants (i-iii), as well as $\langle\omega\rangle$ to $\omega_{\mathrm{peak}}$, indicate that our assumption of a strongly localized peak in the structure factor that is not strongly skewed is sufficiently justified for extracting $\omega_\mathrm{rot}$ from the CL data.

\clearpage

\begin{thebibliography}{78}
\bibitem [{dis()}]{discretization}%
  \BibitemOpen
  \href@noop {} {}\bibinfo {note} {It is not harmful if $\delta_\mathrm{s}$ is
  not large compared to $a$, but within the contact-interaction (s-wave
  scattering) approximation adopted here, it would not make sense to choose
  $\delta_\mathrm{s}$ that small. Since $a$ is typically of the same order of
  magnitude as the extension of the microscopic interaction potential between
  atoms, one would have to insert the precise form of this potential to gain
  further accuracy at this point (which is much harder to determine
  experimentally and theoretically than $a$, which is straightforward to
  measure). Furthermore, if $\delta_\mathrm{s}$ is not large compared to $a$,
  it arises the technical difficulty that one cannot use the perturbative Born
  approximation any more to determine the relationship between $a$ and $g$ as
  outlined in \App{Born_series} (the Born series can be resummed to infinite
  order for a contact interaction, but we are not aware of any such resummation
  for the dipolar case). A numerical approach would thus have to be
  devised.}\BibitemShut {Stop}%
\bibitem [{}]{Langevin_time}%
  \BibitemOpen
  \href@noop {} {}\bibinfo {note} {Note that the Langevin ``time'' has the dimension of a density, as follows from dimensional analysis of the Langevin equations.}\BibitemShut {Stop}%
\end{thebibliography}

%

\end{document}